\documentclass[twoside,11pt]{article}
\usepackage{graphicx}
\usepackage{bm}
\usepackage{float}
\usepackage{tabularray}
\usepackage{booktabs}
\usepackage{lipsum}
\usepackage{cite}
\usepackage{appendix}
\usepackage[margin=1in]{geometry}
\usepackage[usenames]{color}
\usepackage{amsfonts, amssymb, amsmath, amsthm, multicol}
\usepackage[colorlinks=true, pdfstartview=FitV, linkcolor=blue,
citecolor=blue, urlcolor=blue]{hyperref}
\usepackage[usenames]{color}
\definecolor{Red}{rgb}{1,0.05,0}
\definecolor{Grn}{rgb}{0.1,0.7,0.1}
\definecolor{Blu}{rgb}{0.1,0.1,0.6}
\definecolor{Org}{rgb}{1,0.45,0}
\definecolor{Vio}{rgb}{0.6578,0,0.9478}
\definecolor{Mag}{rgb}{1,0.2,0.3}
\usepackage{authblk}
\usepackage{booktabs}
\usepackage{breqn}

\usepackage[section]{placeins}
\usepackage{graphicx}
\usepackage{amssymb}
\usepackage{epstopdf, epsfig}
\usepackage{color}
\usepackage{float}
\usepackage{amsmath}
\usepackage{tabularx}
\usepackage{bm}
\usepackage{accents}
\usepackage[numbers,sort&compress]{natbib}
\usepackage{array}
\usepackage{tabularx}
\usepackage{multirow,multicol, makecell}
\usepackage{pdflscape}
\usepackage{float}
\usepackage{breqn}
\usepackage{mathtools}
\usepackage{breqn}
\usepackage[flushleft]{threeparttable}
\usepackage{booktabs,caption}
\usepackage{footnote}
\usepackage{rotating}
\usepackage{lscape}
\makesavenoteenv{tabular}
\newcolumntype{C}[1]{>{\centering\arraybackslash}p{#1}}
\newcolumntype{L}[1]{>{\raggedright\arraybackslash}p{#1}}
\newcolumntype{M}[1]{>{\centering\arraybackslash}m{#1}}
\usepackage{makecell}

\usepackage[mathlines,displaymath]{lineno}
\usepackage{lscape}
\usepackage{siunitx}
\usepackage{mathtools}
\usepackage[labelfont=bf]{caption}
\usepackage{hyperref}
\hypersetup{
    colorlinks,
    linkcolor={red!50!black},
    citecolor={blue!50!black},
    urlcolor={blue!80!black}
}

\usepackage{adjustbox}
\usepackage{xcolor}
\usepackage{soul}

\numberwithin{rmk}{section}
\numberwithin{nt}{section}

\title{Maximum droplet volume on cylindrical surfaces}

\author[1]{\textcolor{black}{Yi Zhang}}
\author[1]{\textcolor{black}{Apurav Tambe}}
\author[1]{\textcolor{black}{Zhao Pan}\thanks{To whom correspondence may be addressed: {zhao.pan}@uwaterloo.ca}}
\affil[1]{University of Waterloo, Department of Mechanical and Mechatronics Engineering, Waterloo, ON, Canada}

\date{\today}

\begin{document}
\maketitle \vspace{1cm}

\section*{Abstract}

The maximum volume ($\Omega$) of a droplet that can remain attached to a horizontal fiber defines the stability limit of droplet-fiber interactions, phenomena common in nature and critical to diverse engineering applications. 
Existing predictive models for $\Omega$ show limitations in accurately capturing the dependence of $\Omega$ on fiber size and wettability.
To address this gap, we systematically investigate $\Omega$ on a horizontal fiber through numerical simulations and experiments.
A comprehensive semi-empirical model for $\Omega$ is developed and validated against both experimental measurements and reference simulations.
This model establishes a new scaling under which the normalized maximum droplet volume depends solely on the contact angle and remains valid across a wide spectrum from a sub-millimeter thin fiber to the flat-surface limit, regardless of the diverse morphologies that droplets exhibit.

\noindent \textbf{Key words}: Droplet; Horizontal fiber; Maximum volume; Capillary length; Contact angle

\section{Introduction}
\label{sec: introduction}

The largest droplet that can hang from a ceiling before it falls under gravity is a classic manifestation of the Rayleigh-Taylor instability \cite{taylor1950instability}, and has remained an enduring subject of investigation \cite{padday1973stability, daniel2023droplet, sadullah2024predicting}.
Established results reveal that the dimensionless maximum volume of such a droplet ($\bar{\Omega}_*=\bar{\Omega}/\lambda^3$, where $\bar{\Omega}$ is the dimensional maximum volume and $\lambda$ is liquid capillary length) depends solely on the surface contact angle ($\theta$) \cite{padday1973stability}.
Imagine ``rolling up'' the ceiling into a cylinder: the problem of a droplet hanging beneath a plane naturally transforms into that of a droplet hanging on a fiber (despite the markedly altered droplet morphology), a system that has also been extensively studied \cite{carroll1984equilibrium, carroll1986equilibrium, bitten1970coalescence}.

Droplet-fiber interactions are pervasive in both nature and engineering applications, from spider webs capturing water to engineered fibers designed for separation and harvesting devices \cite{zheng2010directional, duprat2012wetting, ju2012multi, park2013optimal}.
Among the many phenomena in this setting, a fundamental question is again the maximum droplet volume ($\Omega$) that a horizontal fiber can sustain, which has critical implications for applications such as fog harvesting \cite{zhang2018highly, hou2012water, mukhopadhyay2024droplet, park2013optimal}, as well as filtration and separation \cite{gu2020study, fang2015drop}. 

To address this deceptively simple question, Lorenceau et al. \cite{lorenceau2004capturing} developed a widely adopted analytical model for droplets on thin fibers (fiber radius $r\ll\lambda$), given by
\begin{equation}
\label{eq:Loren}
\Omega_*=\frac{\Omega}{\lambda^2r} = 4\pi,
\end{equation}
where $\Omega_*$ denotes the dimensionless maximum droplet volume on a fiber.
To account for $\theta$, Wang and Olivier \cite{wang2018numerical} modified the model to $\Omega = 4\pi r\lambda^2 \cos\theta$.
Moreover, Pan et al. \cite{pan2018drop} employed a free-energy analysis and derived $\Omega = 4\pi r\lambda^2 \sin\beta$, where $\beta$ is the droplet off-axis angle influenced by~$\theta$.
However, determining $\beta$ requires experimental input.
These models share the same underlying assumption: the three-phase contact line (TCL) is approximated by two complete circumferences of the fiber cross-section, giving a TCL length of $4\pi r$.
However, as the droplet volume approaches its maximum, it becomes energetically favorable for the droplet to adopt a clam-shell morphology \cite{mchale2002global, chou2011equilibrium}, in which the droplet no longer fully wraps the fiber, and the TCL is not circular but instead becomes irregular. 
This breakdown is pronounced on thick fibers or at low liquid-solid wettability \cite{chou2011equilibrium}, leading to marked discrepancies between theoretical predictions \cite{lorenceau2004capturing, wang2018numerical} and collected data (see Electronic Support Information (ESI) Section 1).
Furthermore, as the fiber thickens and approaches a flat-surface geometry, $\Omega$ should asymptotically converge to the planar ceiling limit $\bar{\Omega}_*\lambda^3$. 
Yet, no existing framework has been able to quantitatively capture this transition.

In this letter, we develop a comprehensive model that spans a wide spectrum from thin fibers to the flat-surface limit, while explicitly incorporating the role of $\theta$.
It resolves the long-standing problem of how droplet retention evolves with fiber radius, bridging the classic ``droplet-on-ceiling'' problem with its cylindrical analogue, ``droplet-on-fiber''.

\section{Methods}
\label{sec: methods}

\subsection{Simulation by surface evolver}
\label{sec: Sim by SE}

Surface Evolver (SE, version 2.70) \cite{brakke1992surface} is employed to investigate the maximum droplet volume~($\Omega$) on a horizontal fiber by simulating the profiles of droplets of progressively increasing volume until detachment occurs.
In SE, the liquid phase is initialized as a cuboid-shaped slab positioned beneath a cylindrical fiber with radius ($r$) and prescribed liquid-solid contact angle ($\theta$), see the simulation setup in ESI Section 2.
The liquid is characterized by its surface tension ($\gamma$), density ($\rho$), and volume ($V$).
SE evolves the liquid cuboid towards its equilibrium shape by minimizing the total free energy ($F$), given by \cite{chou2011equilibrium}
\begin{equation}\label{eq:SE}
F =\gamma\left(A_{LV}-A_{LS}\cos\theta\right) + \iiint\rho gz\mathrm{d}V,
\end{equation}
where $A_{LV}$ and $A_{LS}$ are the liquid-vapor and liquid-solid interfacial area, respectively, and the last term represents the gravitational potential energy. 
The simulation starts with a small $V$ for which the droplet remains stably attached to the fiber ($V \leq \Omega$).
The volume is then increased stepwise by \SI{0.5}{\micro\liter} until detachment occurs ($V > \Omega$), marking the maximum supported volume (see Figure~S2).


In this work, simulations are conducted for $\theta$ ranging from $1^\circ$ to $90^\circ$ and $r$ from 0.1~mm to 9~mm. 
The liquid capillary length, defined as $\lambda =\sqrt{\gamma/\left(\rho g\right)}$, is varied from 1.4~mm to 2.7~mm.
This variation is achieved by adjusting $\gamma$ in the range of \SI{20}{mN/m} to \SI{70}{mN/m}, and $\rho$ from \SI{0.7e3}{kg/\cubic\meter} to \SI{1.00e3}{kg/\cubic\meter}.
The parameters $\theta$, $r$, and $\lambda$ are sampled at a discrete set of representative values, and all combinations within these sets are systematically simulated for $\Omega$.

\subsection{Experimental validation}
\label{sec: Exp valid}

The maximum droplet volumes for various liquid-fiber pairs are measured experimentally.
The test liquid is gradually dispensed onto a horizontally mounted fiber using an electronic pipette until detachment, and the dispensed volume is recorded as $\Omega$.
Table~S1 summaries the tested liquid-fiber pairs and their physical properties.
Six liquids with $\gamma$ ranging from \SI{21}{mN/m} to \SI{72}{mN/m} and $\lambda$ from \SI{1.5}{mm} to \SI{2.7}{mm} are employed.
Fibers of five different materials covering $\theta$ from $0^\circ$ to $70^\circ$ are examined, with $r$ varied between 0.1~mm and 19.1~mm. 
Details of the experimental setup and procedures are provided in ESI Section 3. 
Additionally, experimental and simulation data from the literature \cite{pan2018drop,datta2021droplet,chou2011equilibrium} are incorporated for external validation.

\section{Results and discussion}
\label{sec: Results}

\subsection{Simulation results and semi-empirical model development}
\label{sec: Sim results}

\begin{figure}[!h]
    \centering\includegraphics[width=1\linewidth]{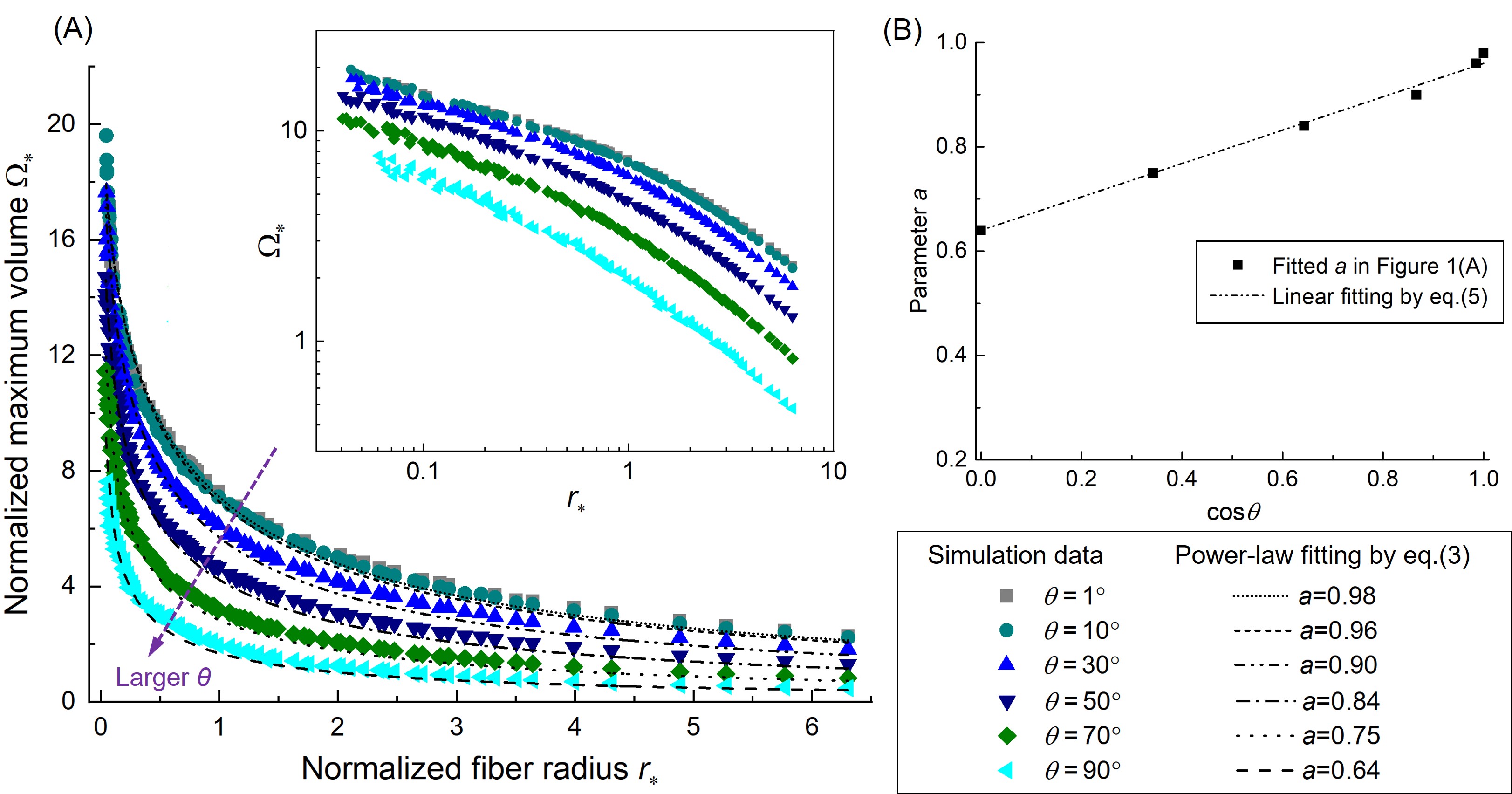}
    \caption{(A) Simulation results of the normalized maximum droplet volume ($\Omega_*=\Omega/\lambda^2r$) as a function of the normalized fiber radius ($r_*=r/\lambda$) for various contact angles ($\theta$).
    Power-law fits of $\Omega_*$ follow eq~\eqref{eq:power-law correlation}.
    (B) Fitted values of parameter $a$ corresponding to (A).}
    \label{fig:sim data}
\end{figure}

Figure~\ref{fig:sim data}(A) presents the SE simulated values of $\Omega_*$ as a function of the normalized fiber radius ($r_* = r/\lambda$) for various contact angles ($\theta$).
In contrast to the constant prediction by Eq~\eqref{eq:Loren}, $\Omega_*$ decreases with increasing $\theta$ and $r_*$.
By rescaling the simulation results onto a logarithmic scale (inset), we observe that the slope of $\Omega_*(r_*,\theta)$ on the log-log plot mostly varies with $r_*$.
This indicates a potential power-law relationship between $\Omega_*$ and $r_*$, characterized by a variable exponent $n$ that captures the slope variation.
Accordingly, we propose the following relationship:
\begin{equation}
\label{eq:power-law correlation}
\Omega_* =\frac{\Omega}{\lambda^2r} \approx \bar{\Omega}_* r_*^{-n},
\end{equation}
where $\bar{\Omega}_*$ denotes the normalized maximum droplet volume under a flat surface.
Although there is no known closed solution for $\bar{\Omega}_*$, numerical results such as the seminal envelope curve (adopted here) developed by Padday and Pitt \cite{padday1973stability}, and several subsequent empirical models \cite{sadullah2024predicting,daniel2023droplet,zhang2025droplets} are available.
We construct the exponent $n$ taking the form
\begin{equation}
\label{eq:n}
n = \frac{a}{r_*-1}+1,
\end{equation}
The parameter, $a$, is determined by fitting eq~\eqref{eq:power-law correlation} to the data in Figure~\ref{fig:sim data}(A). 
The fitted values of $a$ are presented in Figure~\ref{fig:sim data}(B) as a function of $\cos\theta$, revealing a linear dependence given by
\begin{equation}
\label{eq:a}
a \approx 0.32\cos\theta+0.64.
\end{equation}
Eq~\eqref{eq:power-law correlation} can thus be expressed as
\begin{equation}
\label{eq:comp model}
\Omega_* =\frac{\Omega}{\lambda^2r} \approx \bar{\Omega}_* r_*^{-\left(\frac{0.32\cos\theta+0.64}{r_*-1}+1 \right)}.
\end{equation}
In the limiting cases, the asymptotic behaviors of eq~\eqref{eq:comp model} are consistent with physical expectations.
As $r_*\to\infty$ (i.e., when the fiber becomes sufficiently thick to approximate a flat surface), it reduces to $\Omega/\lambda^3 \approx \bar{\Omega}_*$, recovering the classic “droplet-on-ceiling” limit.
As $r_*\to 0$, eq~\eqref{eq:comp model} yields $\Omega\to 0$, as no solid surface exists to support the droplet.
For finite but small $r_*$ ($r_*\ll1$), eq~\eqref{eq:comp model} simplifies to $\Omega/\lambda^{2.04}r^{0.96} \approx \bar{\Omega}_*$ under high wettability, closely resembles the scaling by Lorenceau et al., as relisted in eq~\eqref{eq:Loren} \cite{lorenceau2004capturing}.
This formulation illustrates the evolution of the maximum droplet volume as $r$ increases from a thin fiber to a flat surface, and explicitly incorporates the role of $\theta$.

Another way to interpret eq~\eqref{eq:comp model} is to rearrange the $r_*$-dependent factor to the left-hand side of the equation, leading to a new normalization of $\Omega$, denoted as $\Omega_\dagger$:
\begin{equation}
\label{eq:comp model2}
\Omega_\dagger = \frac{\Omega}{\lambda^3} r_*^{n-1} = C(r_*,\theta) \frac{\Omega}{\lambda^3}  \approx \bar{\Omega}_*(\theta),
\end{equation}
where $C(r_*,\theta) = r_*^{\frac{0.32\cos\theta+0.64}{r_*-1}}$ is a dimensionless coefficient.
Remarkably, $C(r_*,\theta)$ enables a new scaling that eliminates the contribution from the cylinder diameter $r_*$, and the normalized maximum droplet volume on a fiber can be approximated by the flat-surface value $\bar{\Omega}_*$, depending solely on $\theta$.

\subsection{Validation and discussion}
\label{sec: validation}
Figure~\ref{fig:validation}(A) validates the comprehensive model given by eq~\eqref{eq:comp model2} for predicting the maximum droplet volume on a horizontal fiber.
Validation is performed against simulation results from an earlier work \cite{chou2011equilibrium} as well as experimental data obtained both in our laboratory and from the literature \cite{pan2018drop,datta2021droplet}.
Across these datasets, $r_*$ spans a wide range from 0.04 to 12.55, with the upper limit close to a flat surface. 
As an illustration, at $r_*=12.55$, the dimensionless coefficient $C$ in eq~\eqref{eq:comp model2} reaches $\sim1.23$ under perfect wetting, implying that $\Omega$ on such a fiber exceeds 80\% (or 1/1.23) of that under a flat surface.
All $\Omega$ are normalized to $\Omega_\dagger$ following eq~\eqref{eq:comp model2}.
As shown in Figure~\ref{fig:validation}(A), $\Omega_\dagger$ for different $r_*$ and $\theta$ collapse onto a single curve $\bar{\Omega}_*(\theta)$, in contrast to the broad variation of $\Omega_*$ observed in Figures~\ref{fig:sim data} and~S1, thereby validating the relation $C(r_*,\theta) \frac{\Omega}{\lambda^3}  \approx \bar{\Omega}_*(\theta)$ of eq~\eqref{eq:comp model2}.
This collapse is independent of the various droplet morphologies across varying $r_*$ and $\theta$ (see discussion below).
The inset of Figure~\ref{fig:validation}(A) plots $\Omega_\dagger$ versus $\bar{\Omega}_*(\theta)$. 
All data fall reasonably well on the $\Omega_\dagger = \bar{\Omega}_*$ line (dashed), confirming $\Omega_\dagger \approx \bar{\Omega}_*(\theta)$ of eq~\eqref{eq:comp model2}.
Overall, the model agrees well with both the experimental data and the reference simulations, with a mean relative error of 10.0\%.

\begin{figure}[!h]
    \centering
    \includegraphics[width=1\linewidth]{Fig2.jpg}
    \caption{(A) Validation of the developed semi-empirical model, given by eq~\eqref{eq:comp model2}, for predicting the maximum droplet volume on a horizontal fiber. 
    The legend on the right specifies the liquid type, fiber radius, and fiber material. 
    Error bars represent the standard deviation from three to five repeated measurements. 
    (B) Photos of maximum-volume droplets suspended beneath fibers and flat surfaces for $\theta\approx0^\circ$ and $\theta\approx 60^\circ$, corresponding to the marked data points in (A).  
    Scale bar indicates \SI{0.5}{mm}.} 
    \label{fig:validation}
\end{figure}

Figure~\ref{fig:validation}(B) exemplifies the morphologies of maximum-volume droplets suspended beneath various fibers and flat surfaces for $\theta\approx0^\circ$ and $\theta\approx 60^\circ$.
For a given $\theta$, droplets adopt distinct shapes as $r_*$ varies.
The mechanisms governing droplet instability also differ: de-pinning of TCL for thin fibers while necking of droplets for flat surfaces, as discussed in our previous work \cite{zhang2025droplets}.
Despite these contrasting behaviors, the proposed model provides a unified scaling.
With the scaling coefficient $C(r_*,\theta)$, $\Omega_\dagger$ for cylindrical fibers, even across different $r_*$, collapses onto $\bar{\Omega}_*$ for flat surfaces at a given $\theta$, thereby recovering the classic “droplet-on-ceiling” limit.

\begin{figure}[!h]
    \centering
    \includegraphics[width=1\linewidth]{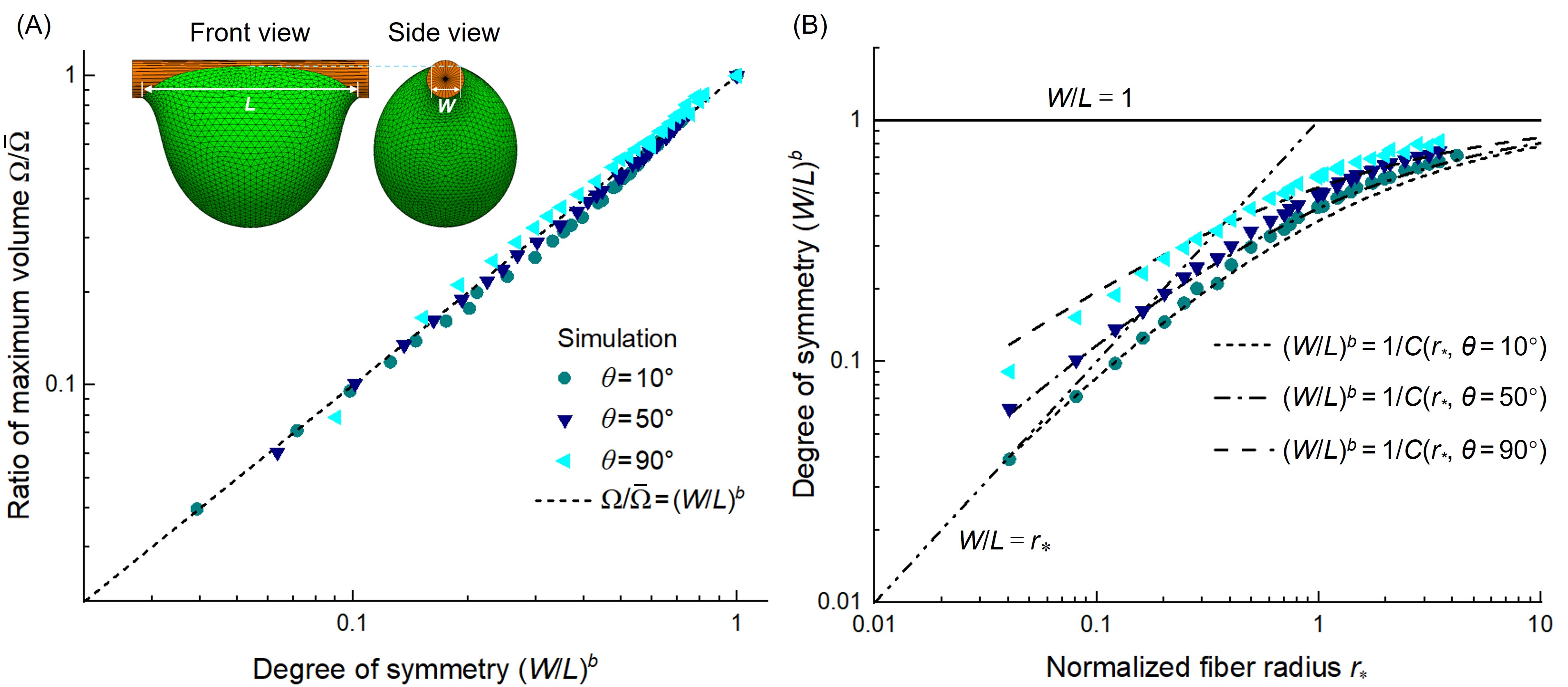}
    \caption{Degree of droplet symmetry, characterized by the ratio of the wetting width $W$ across the fiber to the wetting length $L$ along the fiber. (A) $\Omega/\bar{\Omega}$, the ratio between the maximum droplet volume on a fiber ($\Omega$) and that beneath a flat surface ($\bar{\Omega}$) under the same $\lambda$ and~$\theta$, as a function of $\left(W/L\right)^b$, where $b=0.5\sin\theta+0.9$. (B) $\left(W/L\right)^b$ as a function of $r_*$. Point symbols correspond to those in (A).
    } 
    \label{fig:w/h}
\end{figure}

The physics underlying eq~\eqref{eq:comp model2} may be understood from the asymmetric contact line of a droplet on a fiber.
We characterize the droplet's degree of symmetry by the ratio ($W/L$) between the wetting width $W$ across the fiber and the wetting length $L$ along the fiber (see Figure~\ref{fig:w/h}).
For small $r_*$, the droplet exhibits a low degree of symmetry (small $W/L$), and its maximum volume ($\Omega$) deviates markedly from the flat-surface value ($\bar{\Omega}$).
Increasing $r_*$ enhances the symmetry (larger $W/L$), and $\Omega$ progressively approaches $\bar{\Omega}$.
In the flat-surface limit, symmetry is restored with $W/L=1$ and $\Omega/\bar{\Omega}=1$.
Our simulations reveal (see Figure~\ref{fig:w/h}(A)) a simple scaling relation $\Omega/\bar{\Omega}\approx \left(W/L\right)^b$, where $b=1.00$, 1.27, and 1.43 for $\theta=10^\circ$, $50^\circ$, and $90^\circ$, respectively.
These values are well captured by the fit $b=0.5\sin\theta+0.9$.

Figure~\ref{fig:w/h}(B) further illustrates how $\left(W/L\right)^b$ depends on $r_*$.
Under well-wetting conditions (e.g., $\theta=10^\circ$), for small $r_*$ the lateral ($W$) and axial ($L$) wetting are dominated by $r$ and $\lambda$, respectively, leading to the asymptotic behavior $W/L \to r_*$ as $r_*\to 0$ (dash-dot-dot line).
In the opposite limit ($r_*\to \infty$), the geometry approaches a flat surface and $W/L\to 1$ (solid line).
As an example, the relation $\left(W/L\right)^b = 1/C\left(r_*, \theta=10^\circ\right)$ (short-dashed line) bridges these two limits and shows reasonable agreement with the simulations.
Inserting this relation into the scaling $\Omega/\bar{\Omega} \approx \left(W/L\right)^b$ in Figure~\ref{fig:w/h}(A) recovers eq.~\eqref{eq:comp model2}, i.e., $\Omega/\bar{\Omega} \approx 1/C\left(r_*, \theta\right)$.
For large $\theta$, the asymptotic scaling $W/L \to r_*$ as $r_*\to 0$ no longer holds, as $\theta$ significantly suppresses droplet spreading along the fiber $L$.
Nevertheless, the expressions $\left(W/L\right)^b = 1/C\left(r_*, \theta=50^\circ\right)$ and $\left(W/L\right)^b = 1/C\left(r_*, \theta=90^\circ\right)$ remain in reasonable agreement with the simulations for $\theta=50^\circ$ and $\theta=90^\circ$, respectively.


\section{Conclusion}
\label{sec: Conclusion}
In summary, this work systematically investigates the ability of horizontal fibers to retain liquid droplets.
We develop and validate a unified semi-empirical model for the maximum droplet volume ($\Omega$), expressed as $C \frac{\Omega}{\lambda^3} \approx \bar{\Omega}_*$, where $C$ is a dimensionless coefficient dependent on the normalized fiber radius and contact angle; $\lambda$ is the capillary length; $\bar{\Omega}_*$ is the normalized maximum droplet volume under the corresponding flat surface.
The model remains valid across the wide transition from a sub-millimeter thin fiber to the flat-surface limit.
We further show that $\Omega$ is governed by the degree of symmetry of the droplet, as it becomes asymmetric when wetting a cylindrical fiber.

\section*{Acknowledgement}
The authors gratefully acknowledge the financial support provided by the Artificial Intelligence for Design Challenge (AI4D) program, funded by the National Research Council Canada. 

\bibliographystyle{unsrt}

\bibliography{reference}

\end{document}